\documentclass{ws-procs9x6}
\newcommand{\VEV}[1]{\langle #1 \rangle}

\begin{document}

\title{Chromomagnetic Instability and Gluonic Phase\footnote{
Talk given at {\it
2006 International Workshop on ``Origin of Mass and
Strong Coupling Gauge Theories (SCGT06)''},
Nagoya, Japan, November 21-24, 2006.}
}

\author{Michio Hashimoto}

\address{Department of Physics, Nagoya University, \\
Nagoya, 464-8602, Japan\\
E-mail: michioh@eken.phys.nagoya-u.ac.jp}

\begin{abstract}

We briefly report on a recent development in studies of a phase with
vector condensates of gluons (gluonic phase) in dense two-flavor 
quark matter.
\end{abstract}


\bodymatter

\section{Introduction}
\label{sec1}

It has been suggested that quark matter might exist inside 
central regions of compact stars.\cite{quark_star}
At sufficiently high baryon density,
cold quark matter is expected to be in a color superconducting (CSC) 
state.\cite{CSC}
This is one of the reasons why the color superconductivity has been
intensively studied.\cite{review}

Bulk matter in the compact stars must be in $\beta$-equilibrium 
and be electrically and color neutral.
The electric and color neutrality conditions play a crucial role 
in the dynamics of 
quark pairing.\cite{neutrality}
In addition, the strange quark mass cannot be neglected
in moderately dense quark matter.
Then a mismatch $\delta\mu$ between the Fermi surfaces of
the pairing quarks is induced.

As the mismatch $\delta\mu$ grows,
the CSC state tends to be destroyed.
However, the dynamics is not yet solved completely.
It is one of the central issues in this field 
to reveal the phase structure.

The problem is that the (gapped/gapless) two-flavor color 
superconducting phase (2SC/g2SC) suffers from 
the chromomagnetic instability connected with 
tachyonic Meissner screening masses of gluons.\cite{Huang:2004bg}
While the Meissner mass for the 8th gluon is imaginary 
in the g2SC phase $\delta\mu > \Delta$, 
where $\Delta$ is a diquark gap,
the chromomagnetic instability for the 4-7th gluons appears 
in $\delta\mu > \Delta/\sqrt{2}$.
Such a chromomagnetic instability has been found also in 
the gapless color-flavor locked (gCFL) 
phase.\cite{Casalbuoni:2004tb,Alford:2005qw,Fukushima:2005cm}

Since the Meissner masses are defined at zero momentum and 
thus {\it not} pole ones,
the physical origin of the chromomagnetic instability
is not obvious.
We then study the spectrum of plasmons with 
nonzero energy and momenta.\cite{Gorbar:2006up}
We also analyze the dispersion relations of
the diquark fields.\cite{Hashimoto:2006mn}
Then we will find that certain instabilities appear both in 
$\delta\mu > \Delta/\sqrt{2}$ and $\delta\mu > \Delta$, 
corresponding to the chromomagnetic instability.

How can we resolve the problem? 
Actually, numbers of ideas have been proposed by several 
authors.\cite{Alford:2000ze,Reddy:2004my,Huang:2005pv,Hong:2005jv,Kryjevski:2005qq,Gorbar:2005rx}

As a candidate of the genuine vacuum, 
we introduce a phase with vectorial 
gluon condensates (gluonic phase).\cite{Gorbar:2005rx} 
We also show that the single plane wave 
Larkin-Ovchinnikov-Fulde-Ferrell (LOFF) 
phase\cite{LOFF,Alford:2000ze,Giannakis:2004pf}
suffers from a chromomagnetic instability.\cite{Gorbar:2005tx} 

\section{Spectra of the plasmons and the diquark excitations}

\begin{figure}[tbp]
   \begin{center}
     \resizebox{0.45\textwidth}{!}{\includegraphics{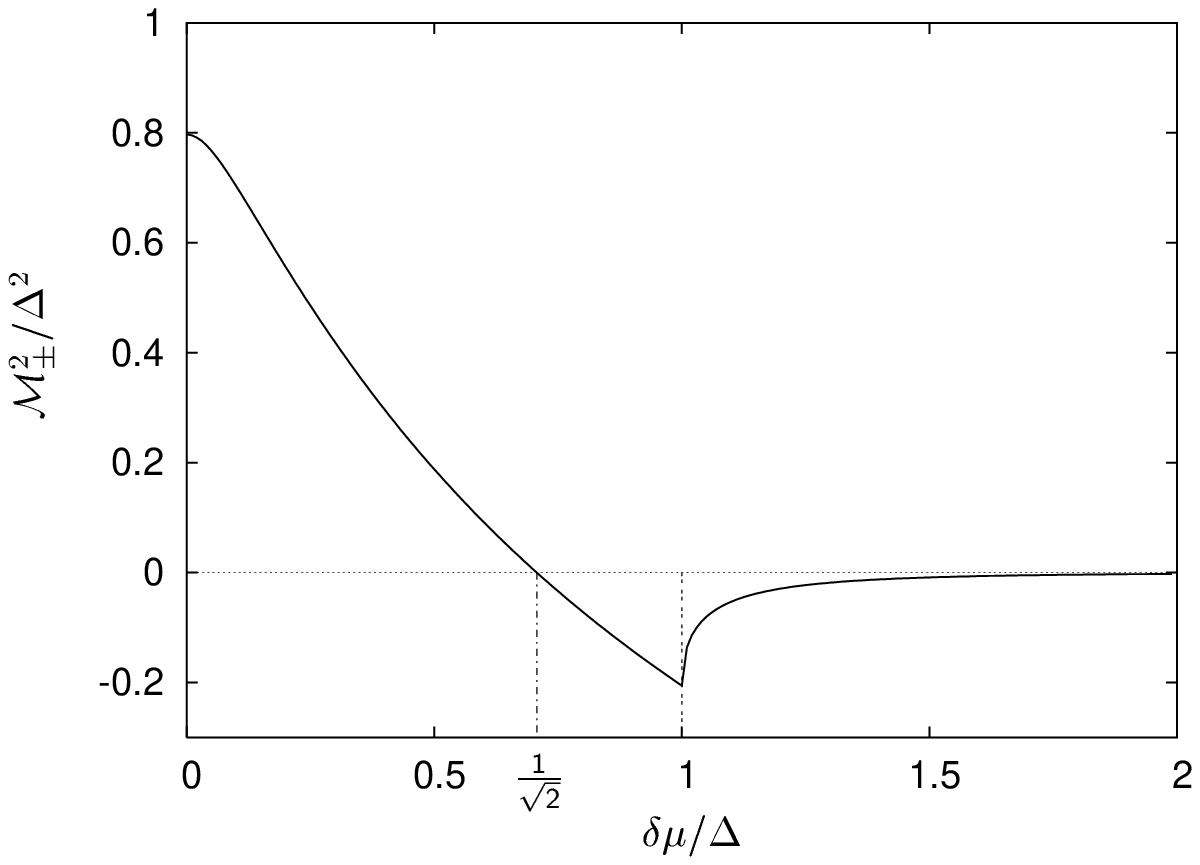}}\qquad
     \resizebox{0.45\textwidth}{!}{\includegraphics{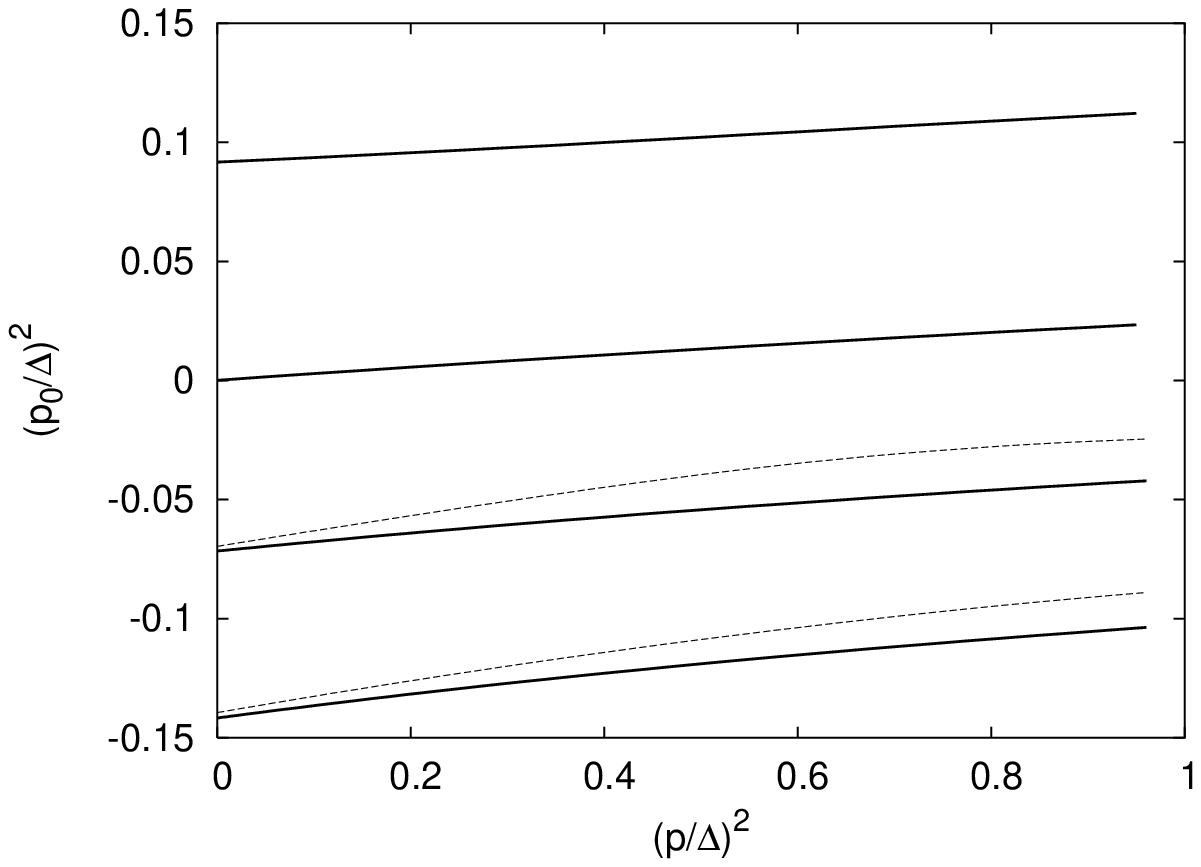}}
   \end{center}
\caption{Mass-gap-squared of the magnetic and electric plasmons 
         for the 4-7th gluons (l.h.s.) and
         dispersion relations for the 4-7th magnetic plasmons (r.h.s.). 
         In the r.h.s., the bold lines are for 
         $\delta\mu/\Delta = 0.6, 1/\sqrt{2}$, $0.8$, and $0.9$, 
         (2SC region) from top to bottom, while the dashed lines are for
         $\delta\mu/\Delta = 1.065$ and $1.009$, 
         (g2SC region) from top to bottom.
\label{fig1}}
\end{figure}

Let us study two point functions of gluons and the diquark fields.

First, we analyze the polarization functions of gluons.\cite{Gorbar:2006up}

For the 4-7th gluons, we find that the mass gaps for magnetic and 
electric modes coincide. 
We depict the results in the l.h.s. of Fig.~\ref{fig1}.
In the 4-7th channel, there exist the light plasmons 
in the whole region of $\delta\mu/\Delta$.
While in $\delta\mu < \Delta/\sqrt{2}$ 
the light plasmons have the positive mass-gap-squared, 
$0 < {\cal M}_\pm^2 \lesssim \Delta^2$,
in $\delta\mu > \Delta/\sqrt{2}$ the plasmons become tachyons 
with ${\cal M}_\pm^2 < 0$. 
It is noticeable that the instability occurs {\it both} for 
the magnetic and electric modes. 
This is essentially different from the chromomagnetic instability 
where the Debye mass for the electric mode remains real.\cite{Huang:2004bg} 

The dispersion relations for the magnetic and electric modes 
do not coincide in general. 
We plot the dispersion relations of the magnetic modes for 
the 4-7th channels in the r.h.s. of Fig.~\ref{fig1}.
The dispersion relations qualitatively read 
$p_0^2=m^2+v^2p^2$, $(p \equiv |\vec p|)$ 
with $m^2 > 0$ for $\delta\mu/\Delta<1/\sqrt{2}$ 
and $m^2 < 0$ for $\delta\mu/\Delta>1/\sqrt{2}$.
The velocity $v$ is always real and less than 1. 

\begin{figure}[tbp]
   \begin{center}
     \resizebox{0.45\textwidth}{!}{\includegraphics{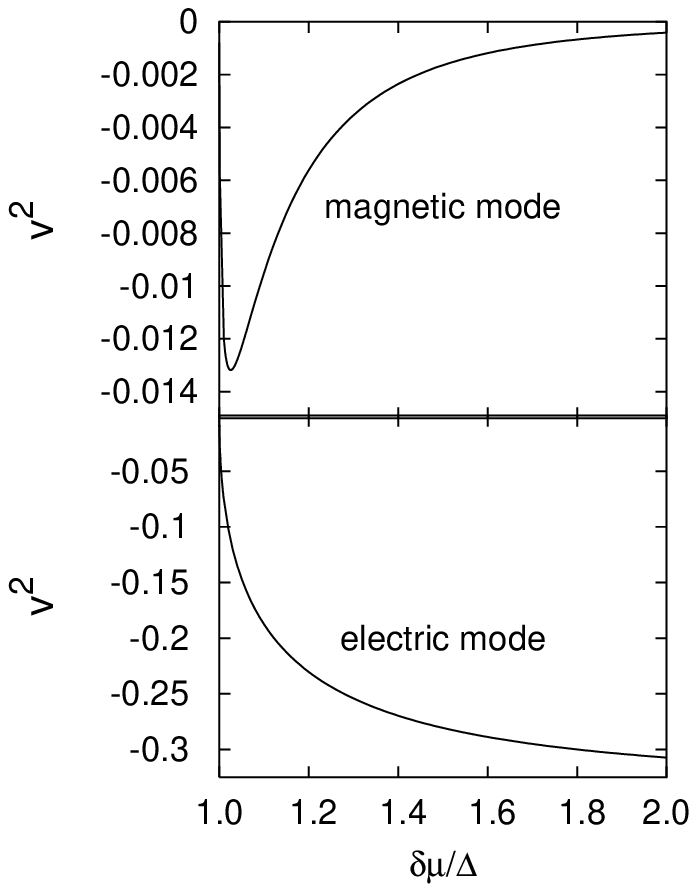}}
     \qquad
     \resizebox{0.45\textwidth}{!}{\includegraphics{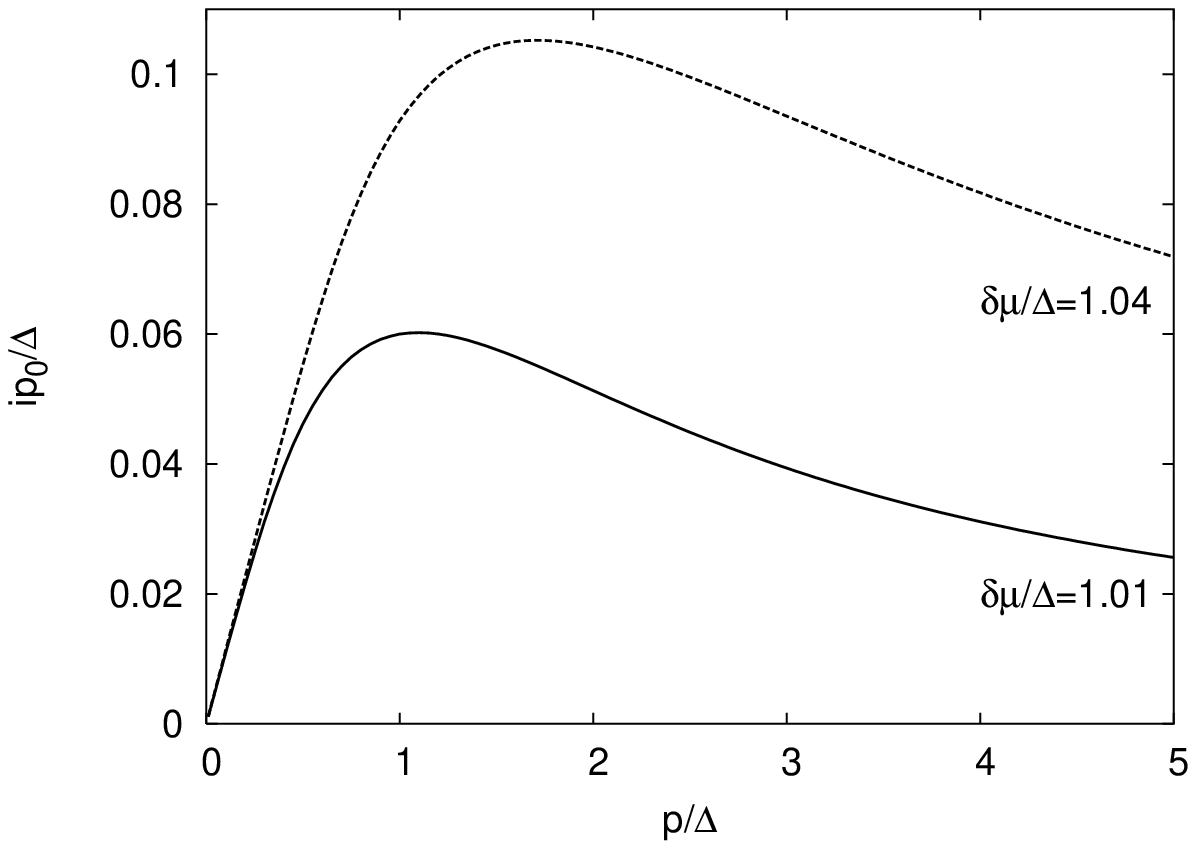}}
   \end{center}
\caption{Velocity squared for gapless tachyons in 
the magnetic and electric modes of the 8th channel (l.h.s.) and
dispersion relations for the gapless magnetic tachyon (r.h.s.).
\label{fig2}}
\end{figure}

For the 8th channel, there is no light plasmon 
in $\delta\mu < \Delta$. 
On the other hand, in the g2SC region $\delta\mu > \Delta$ 
there appear gapless tachyons 
having the dispersion relation
$p_{0}^2 = v^{2} p^2$, $(p \simeq 0)$ with $v^2 < 0$,
as shown in the l.h.s. of Fig.~\ref{fig2}.
This instability occurs {\it both} for the magnetic and electric modes, 
so that it also differs from the chromomagnetic instability 
in the 8th channel.\cite{Huang:2004bg} 
For the dispersion relations of the magnetic mode, 
see the r.h.s. of Fig.~\ref{fig2}.

The appearance of the gapless tachyons in the 8th channel seems 
to be counter-intuitive,
because the Meissner mass squared is nonzero and negative 
in the g2SC phase.\cite{Huang:2004bg}
The origin is a singular dependence on $p_0/p$ in 
the polarization function of the 8th gluon.\cite{Gorbar:2006up}

Let us turn to discuss the two point functions 
of the diquark fields
in the framework of a Nambu-Jona-Lasinio (NJL) model,\footnote{
As usual, we take the anti-blue direction for 
the vacuum expectation value (VEV) of the diquark fields $\Phi^\alpha$, 
$(\alpha=r,g,b)$, i.e., $\VEV{\Phi^b}=\Delta$.
If the model is gauged,
the anti-red and green diquark fields $\Phi^{r,g}$ are eaten by 
the 4-7th gluons, while the imaginary part $\phi_{b2}$ of 
the anti-blue diquark field $\Phi^b$ becomes
the longitudinal mode of the 8th gluon.
Only the real part $\phi_{b1}$ is left as a physical mode.
} based on Ref.~\refcite{Hashimoto:2006mn}. 

While we omit the results for $\Phi^{r,g}$, we do not those 
for $\phi_{b2}$ for clarity.

\begin{figure}[tbp]
  \begin{center}
    \resizebox{0.5\textwidth}{!}{\includegraphics{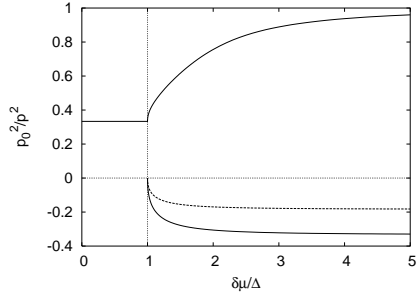}}
  \end{center}
\caption{Velocity squared for the anti-blue diquark field $\Phi^b$.
         The bold dashed and solid curves are for
         $\phi_{b1}$ and $\phi_{b2}$, respectively, where
         $\phi_{b1}$ and $\phi_{b2}$ are the real and imaginary parts
         of $\Phi^b$, respectively.
         We ignored heavy excitations for $\phi_{b1}$.
\label{fig3}}
\end{figure}

We depict the dispersion relations for $\phi_{b1}$ (dashed curve) 
and $\phi_{b2}$ (bold curves) in Fig.~\ref{fig3}.
We find that the velocity squared for $\phi_{b2}$ is 
$v^2=1/3$ in $\delta\mu < \Delta$, 
while there are two branches in $\delta\mu > \Delta$; 
one is, say, an ultra-relativistic branch with $1/3 < v^2 < 1$ and 
the other is a tachyonic one with $v^2 < 0$.
For $\phi_{b1}$, we ignore a heavy mode,
\cite{Ebert:2004dr,He:2005mp,Ebert:2006bq} 
because it should be irrelevant to any instabilities.
Actually, there is no light excitation of $\phi_{b1}$
in $\delta\mu < \Delta$.
Surprising is that even for $\phi_{b1}$ 
a gapless tachyon with $v^2 < 0$ emerges in 
$\delta\mu > \Delta$,\footnote{
A similar instability is also discussed in  
Refs.~\refcite{Iida:2006df,Giannakis:2006gg}.
}
as shown in the dashed curve in Fig.~\ref{fig3}.
(Another branch for $\phi_{b1}$ corresponding to a heavy mode may 
exist in $\delta\mu > \Delta$.\cite{Ebert:2006bq})

We here comment that a singular dependence on $p_0/p$ in 
the two point function of $\Phi^b$ causes
the peculiar behaviors of the dispersion relations for $\phi_{b1}$ and 
$\phi_{b2}$ in $\delta\mu > \Delta$.\cite{Hashimoto:2006mn} 

\section{Gluonic phase}

What do the instabilities imply?
Our answer is the existence of
the vectorial gluon condensates (gluonic phase). 

In order to clarify the problem, let us consider 
the $SU(2)_c$ decomposition owing to the symmetry breaking 
$SU(3)_c \to SU(2)_c$ in the presence of 
the diquark condensate $\Delta$.
The adjoint representation of $SU(3)_c$, i.e., the gluon field $A_\mu^a$, 
$(a=1,2,\cdots,8)$, is decomposed into 
${\bf 3}\oplus{\bf 2}\oplus\bar{{\bf 2}}\oplus{\bf 1}$, namely,
$\{A_\mu^{a}\} = (A_\mu^{1},A_\mu^{2},A_\mu^{3})
 \oplus \phi_\mu \oplus \phi_\mu^* \oplus A_\mu^{8}$.
Here we introduced the complex $SU(2)_c$ doublets of 
the vectorial ``matter'' fields,
\begin{equation}
  \phi_\mu \equiv
  \left(
  \begin{array}{@{}c@{}} \phi_\mu^r \\[2mm] \phi_\mu^g \end{array}
  \right)
  =
 \frac{1}{\sqrt{2}}
  \left(
  \begin{array}{c}
  A_\mu^{4}-iA_\mu^{5} \\[2mm] A_\mu^{6}-iA_\mu^{7}
  \end{array}
  \right) , \qquad
  \phi_\mu^* \equiv
  \left(
  \begin{array}{@{}c@{}} \phi_\mu^{*r} \\[2mm] \phi_\mu^{*g} \end{array}
  \right) .
\label{phi}
\end{equation}
In the gluonic phase, the spatial component of $\phi_\mu$ 
develops the VEV, so that
the role is quite similar to the Higgs doublet in the standard model.

Let us assume nonvanishing VEVs as
\begin{equation}
  B \equiv g\VEV{A^{6}_{z}}, \quad
  C \equiv g\VEV{A^{1}_{z}}, \quad
  D \equiv \mu_3 = g\VEV{A^{3}_{0}}, 
  \label{def-BCD}
\end{equation}
where $C$ is required for consistency with the Ginzburg-Landau (GL) 
approach which we will employ. 
For a more general ansatz, see Ref.\refcite{Gorbar:2007vx}.

Neglecting irrelevant terms, 
we obtain a reduced GL potential,\cite{Gorbar:2005rx}
\begin{equation}
  \tilde{V}_{\rm eff} =
  V_\Delta + \frac{1}{2}M_B^2 B^2
  + T D B^2 + \frac{1}{2}\lambda_{BC} B^2 C^2
  + \frac{1}{2}\lambda_{CD} C^2 D^2 ,
  \label{V_min}
\end{equation}
where $V_\Delta$ is the 2SC part and
the parameter $M_B^2$ is expressed through 
the Meissner mass. 
The negative $M_B^2$ essentially dictates a nonvanishing 
gluon condensate $B \ne 0$.
The coefficients $T$, $\lambda_{BC}$ and $\lambda_{CD}$ are
calculated in the fermion one-loop approximation.\cite{Gorbar:2005rx}
We then find that $\lambda_{CD}$ is definitely negative
and $\lambda_{BC} > 0$ in the vicinity of the critical point 
$\delta\mu \approx \Delta/\sqrt{2}$.
The free energy at the stationary point is found as
\begin{equation}
  \tilde{V}_{\rm eff} =
  V_\Delta - \frac{(-M_B^2)^3}{54 T^2}
  \left(\,-\frac{\lambda_{CD}}{\lambda_{BC}}\,\right) < V_\Delta .
\end{equation}
Therefore the gluonic vacuum is stabler than the 2SC one.
We can check also that the solution corresponds to a minimum. 

It is noticeable that the above gluonic solution describes 
{\it non-Abelian} constant chromoelectric fields, $F_{0j}^{a} \ne 0$.
In this sense, the gluonic phase enjoys a non-Abelian nature.
We emphasize that while an Abelian constant electric field 
always leads to an instability, 
non-Abelian one does not in many cases.\cite{Brown:1979bv}
The difference seems to be connected with the fact that 
a constant Abelian electric field is derived only from 
a vector potential depending on spatial and/or time coordinates,
while a constant non-Abelian chromoelectric field can be expressed 
through constant vector potentials owing to nonzero commutators.
Thus energy and momentum can be left as good quantum numbers 
in the non-Abelian case.

In the gluonic phase, both the rotational $SO(3)$ and 
the electromagnetic $U(1)$ symmetries are spontaneously broken down. 
Therefore, this phase describes an anisotropic medium with 
the color and electromagnetic Meissner effects.
There also exist exotic hadrons in the medium.

\section{The single plane wave LOFF state and its instability}

\begin{figure}[tbp]
  \begin{center}
    \resizebox{0.45\textwidth}{!}{\includegraphics{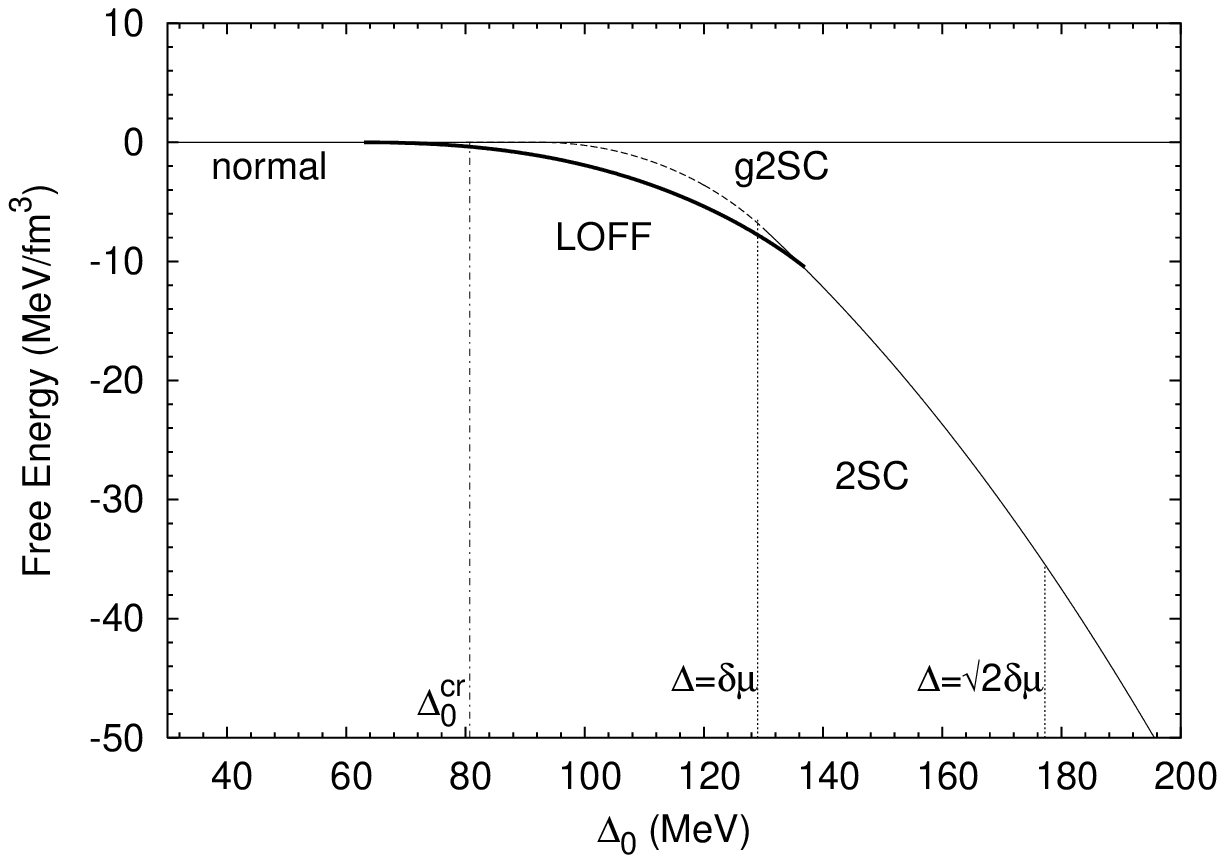}}
    \qquad
    \resizebox{0.45\textwidth}{!}{\includegraphics{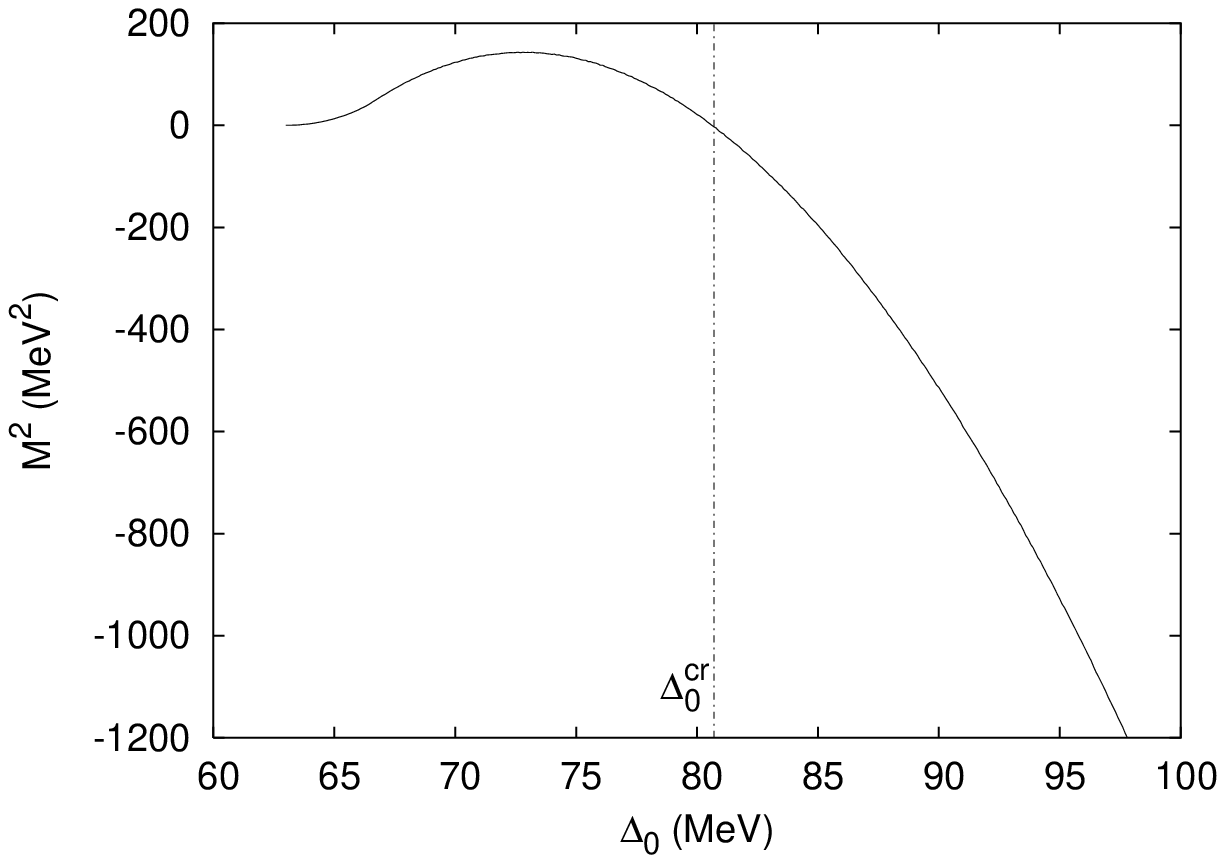}}
  \end{center}
\caption{Free energy of the neutral LOFF state [bold line], and
        the neutral 2SC/g2SC state [solid/dashed line] (l.h.s.)
        and Meissner mass squared $M^2$ for the 4-7th gluons 
        in the LOFF phase (r.h.s.).
        In the l.h.s. the free energy of the neutral normal state 
        is chosen as a reference point. 
        In the r.h.s. we took the QCD coupling constant $\alpha_s = 1$.
\label{energy_diff}}
\end{figure}

In order to demonstrate how the neutrality conditions work and 
dramatically change the situation, we analyze the single plane wave 
LOFF state\footnote{
The order parameter of the single plane wave LOFF state 
has the form
$\VEV{\Phi^b(x)} = \Delta e^{2i\vec q \cdot \vec x}$ 
with a constant phase vector $\vec{q}$, instead of 
$\VEV{\Phi^b} = \Delta$ in the 2SC/g2SC phase.
This phase is gauge equivalent to the phase with 
$\VEV{\Phi^b} = \Delta$ and $\VEV{\vec A^{8}} \ne 0$.
\cite{Gorbar:2005rx,Gorbar:2005tx}
}.\cite{Gorbar:2005tx} 

For a numerical analysis, we fix the quark chemical potential 
$\mu$=400 MeV and the cutoff $\Lambda$ = 653.3 MeV,
and vary the value of $\Delta_0$, which is the 2SC gap parameter 
at $\delta \mu=0$.
We show the free energy differences in the l.h.s. of
Fig.~\ref{energy_diff} (the reference point is the free energy of 
the normal phase with $\Delta=0$). 
The results are not sensitive to the choice of $\mu \;(=\mbox{300--500 MeV})$.

One can see that the neutral LOFF phase is energetically stabler than 
the neutral normal phase and the neutral g2SC/2SC one
in the whole region of the g2SC plus a narrow region 
of the 2SC near the edge, i.e.,
$\mbox{63 MeV} < \Delta_0 < \mbox{136 MeV}$.
Since the chromomagnetic instability in the 2SC phase
occurs in the region $\Delta < \sqrt{2}\delta\mu$, 
which corresponds to $\Delta_0=$ 177 MeV in the l.h.s. of 
Fig.~\ref{energy_diff}, 
the neutral LOFF solution cannot cure the instability.

Furthermore, by applying the Meissner mass formula 
in the second paper in Ref.\refcite{Giannakis:2004pf} 
to our LOFF solution, we find that
the neutral LOFF state
itself suffers from a chromomagnetic instability in 
$\Delta_0 > \Delta_0^{\rm cr}=$ 81 MeV and 
thereby it should not be the genuine ground state.
(See the r.h.s. of Fig.~\ref{energy_diff}.)

\section{Summary and discussions}

We showed that in the 2SC/g2SC phase there appear several instabilities 
other than the chromomagnetic one.

We introduced the gluonic phase to resolve the instabilities.
We also found that the neutral LOFF state is not free from 
the instabilities. 
It indicates that the gluonic phase is relevant 
for curing the problem.\footnote{
See also a numerical approach of the gluonic 
phase.\cite{Kiriyama:2006ui}
}

It is worthwhile to examine 
the multiple plane-wave 2SC LOFF state.\cite{Bowers:2002xr}
It would be also interesting to figure out whether or not 
a phase with vectorial gluon condensates exists in three-flavor
quark matter.

\end{document}